\documentclass[sn-mathphys]{sn-jnl}

\usepackage{breqn}

\jyear{2022}%

\theoremstyle{thmstyleone}%
\theoremstyle{thmstyletwo}%

\theoremstyle{thmstylethree}%

\begin{document}

\title[Deep-learning-based reconstruction of 3D flows from 2D data]{A deep-learning approach for reconstructing 3D turbulent flows from 2D observation data}


\author[1]{\fnm{Mustafa Z.} \sur{Yousif}}\email{mustafa\textunderscore yousif@pusan.ac.kr}

\author[1]{\fnm{Linqi} \sur{Yu}}\email{linqi\textunderscore yu@pusan.ac.kr}

\author[2]{\fnm{Sergio} \sur{Hoyas}}\email{serhocal@mot.upv.es}

\author[3]{\fnm{Ricardo} \sur{Vinuesa}}\email{rvinuesa@mech.kth.se}

\author*[1]{\fnm{HeeChang} \sur{Lim}}\email{hclim@pusan.ac.kr}

\affil*[1]{\orgdiv{School of Mechanical Engineering}, \orgname{Pusan National University}, \orgaddress{\street{Busandaehak-ro}, \city{Busan}, \postcode{46241}, \country{Rep. of KOREA}}}

\affil[2]{\orgdiv{Instituto Universitario de Matemática Pura y Aplicada}, \orgname{Universitat Politècnica de València}\orgaddress{\street{}, \city{València}, \postcode{46022}, \country{Spain}}}

\affil[3]{\orgdiv{FLOW, Engineering Mechanics}, \orgname{KTH Royal Institute of Technology}\orgaddress{\street{}, \city{Stockholm}, \postcode{10044}, \country{Sweden}}}


\abstract{Turbulence is a complex phenomenon that has a chaotic nature with multiple spatio-temporal scales, making predictions of turbulent flows a challenging topic.  Nowadays, an abundance of high-fidelity databases can be generated by experimental measurements and numerical simulations, but obtaining such accurate data in full-scale applications is currently not possible. This motivates utilising deep learning on subsets of the available data to reduce the required cost of reconstructing the full flow in such full-scale applications. Here, we develop a generative-adversarial-network (GAN)-based model to reconstruct the three-dimensional velocity fields from flow data represented by a cross-plane of unpaired two-dimensional velocity observations. The model could successfully reconstruct the flow fields with accurate flow structures, statistics and spectra. The results indicate that our model can be successfully utilised for reconstructing three-dimensional flows from two-dimensional experimental measurements. Consequently, a remarkable reduction of the experimental setup cost can be achieved.}




\maketitle
\section*{Introduction}\label{sec1}

With the development of computational power and experimental tools, an accurate description of various types of turbulent flows can be achieved. In terms of numerical simulations, direct numerical simulation (DNS) can resolve the flow structures from the large energy-containing scales to the smallest scales in the dissipation range \citep{Kolmogorov1941}. In the area of experimental fluid dynamics, significant technical advances have been achieved through particle-image velocimetry (PIV) \citep{Adrian1984}, tomographic-PIV (tomo-PIV) \citep{Scarano2012} and four-dimensional time particle-tracking velocimetry (4D-PTV) \citep{Schanzetal2016}. However, in both the numerical and experimental approaches, substantial costs are required to accurately describe the physics of the turbulent flows and these costs are proportional to the Reynolds number. This nondimensional parameter is proportional to the size of the problem, its characteristic velocity, and inversely proportional to the fluid kinematic viscosity.\par

On the other hand, an enormous amount of data can be generated from experimental and numerical studies. Turbulence is now a science that needs, more than ever, new questions more than new data to solve those questions. This motivates developing data-driven methods that can practically utilise the data for addressing various turbulence-related problems. With the recent rapid advances of machine-learning algorithms and the remarkable improvement in the graphic-processing-unit (GPU) capabilities, machine learning has been applied in various fields, including image processing, natural language processing, robotics, weather forecasting, etc. In terms of fluid dynamics, deep-learning algorithms have been effectively applied to tackle a wide range of problems \citep{Bruntonetal2020, Kutz2017, Vinuesa&Brunton2022}, where deep learning is a subset of machine learning in which neural networks with multiple layers are utilised in the model \citep{LeCunetal2015}. Unlike the conventional linear methods, deep-learning-based techniques can deal with complex non-linear problems. This makes the deep-learning approach a good candidate to be applied for various problems in turbulence, such as turbulence modelling \citep{Duraisamyetal2019, Gamahara&Hattori2017, Lingetal2016}, prediction of turbulent flows \citep{Lee&You2019, Srinivasanetal2019, Yousifetal2022a}, reduced-order modelling \citep{Nakamuraetal2021, Yousif&Lim2022}, turbulent-flow control \citep{Fanetal2020, Han&Huang2020, Rabaultetal2019, Vinuesa&Brunton2022} and non-intrusive sensing \citep{Guastonietal2021, Guemesetal2021}. The reconstruction of turbulent flows from spatially-limited data using deep-learning-based models has been recently a topic of interest, considering the ability of several deep-learning models to map the flow fields that are represented by spatially limited or low-resolution data to high-resolution flow fields \citep{Fukamietal2019, Fukamietal2021, Guemesetal2021, Kimetal2021, Liuetal2020, Yousifetal2021, Yousifetal2022b}. The results obtained from the studies cited above indicate that several deep-learning models have a remarkable potential to map turbulent flow fields with spatially limited distribution to high fidelity flow fields by making use of the available training data. Thus, such deep-learning models possibly can reconstruct the missing regions in the flow fields by compensating them with approximation functions represented by the trainable parameters in the models. \par

In this study, we present a deep-learning-based method to reconstruct three-dimensional (3D) turbulent flows from two-dimensional (2D) data, in such a way that mimics the reconstruction of 3D turbulent flows from 2D PIV measurements. In contrast with the proposed studies in the literature, which are based on simple approaches such as assumptions of frozen velocity \citep{Bruckeretal2012, Brucker1995, Zhangetal2008} and frozen turbulence via Taylor hypothesis \citep{Ganapathisubramanietal2008}, exploiting homogeneity in the flow \citep{Chandramoulietal2019} and proper-orthogonal decomposition (POD) \citep{Braudetal2004, Hamdietal2018}, we apply a deep-learning-based approach to map unpaired intersected 2D turbulent flow sections to 3D flow fields. We propose a generative-adversarial-network (GAN)-based model, 2D3DGAN, to reconstruct three-dimensional turbulent flows. Unlike the traditional convolutional-neural-networks (CNNs)-based models, GAN-based models \citep{Goodfellowetal2014} have shown the ability to capture high-frequency data in detail, and remarkable accuracy in terms of image transformation and super-resolution problems \citep{Ledigetal2017, Mirza&Osindero2014, Wangetal2018, Zhuetal2017}. In a typical GAN, two networks, namely the generator ($G$) and the discriminator ($D$), compete with each other. Here, $G$ generates artificial samples similar to the real ones, whereas $D$ distinguishes the artificial samples from the real ones. The goal of the training process is to make $G$ generate artificial samples that are difficult to distinguish using $D$. We utilise a combination of 2D and 3D CNNs to build the 2D3DGAN, which can make use of both the supervised deep-learning method and the adversarial networks, $i.e.$ $G$ and $D$ networks. We remark that the 2D3DGAN is robust to increasing Reynolds number and the complexity of the flow.\par

\section*{Results}\label{sec2}

\textbf{Building 2D3DGAN architecture.} The process of training GAN can be expressed as a min-max two-player game with a value function $V(D, G)$ such that:

\begin{equation} \label{eqn:eq1}
\begin{split}
\substack{\rm min\\{G}} ~\substack{\rm max\\{D}} ~V(D,G) = \mathbb{E}_{x_r \sim P_{\rm data}(x_r)} [ {\rm log} D(x_r )] + \mathbb{E}_{z \sim P_z(z) } [ {\rm log} (1-D(G(z)))],
\end{split}
\end{equation}

\noindent where $x_r$ represents real data and $P_{\rm data} (x_r)$ is its distribution. Note that $\mathbb{E}$ represents the operation of calculating the average of all the data in the training mini-batch. In the second term of the right-hand side of equation~(\ref{eqn:eq1}), $z$ is a random vector used as an input to $G$, whereas $D(x_r)$ represents the probability that the data is real and not artificial. The output from $G$, $i.e.$ $G(z)$, is expected to generate data that is similar to the real one, such that the value of $D(G(z))$ is close to 1. On the other hand, in $D$, $D(x_r )$ returns a value close to 1, whereas $D(G(z))$ returns a value close to 0. Thus, $G$ is trained in a direction that minimizes $V(D,G)$, and $D$ is trained in a direction that maximizes $V(D,G)$. Additional details are provided in the Methods section. \par

\begin{figure}
  \centerline{\includegraphics[scale=0.16]{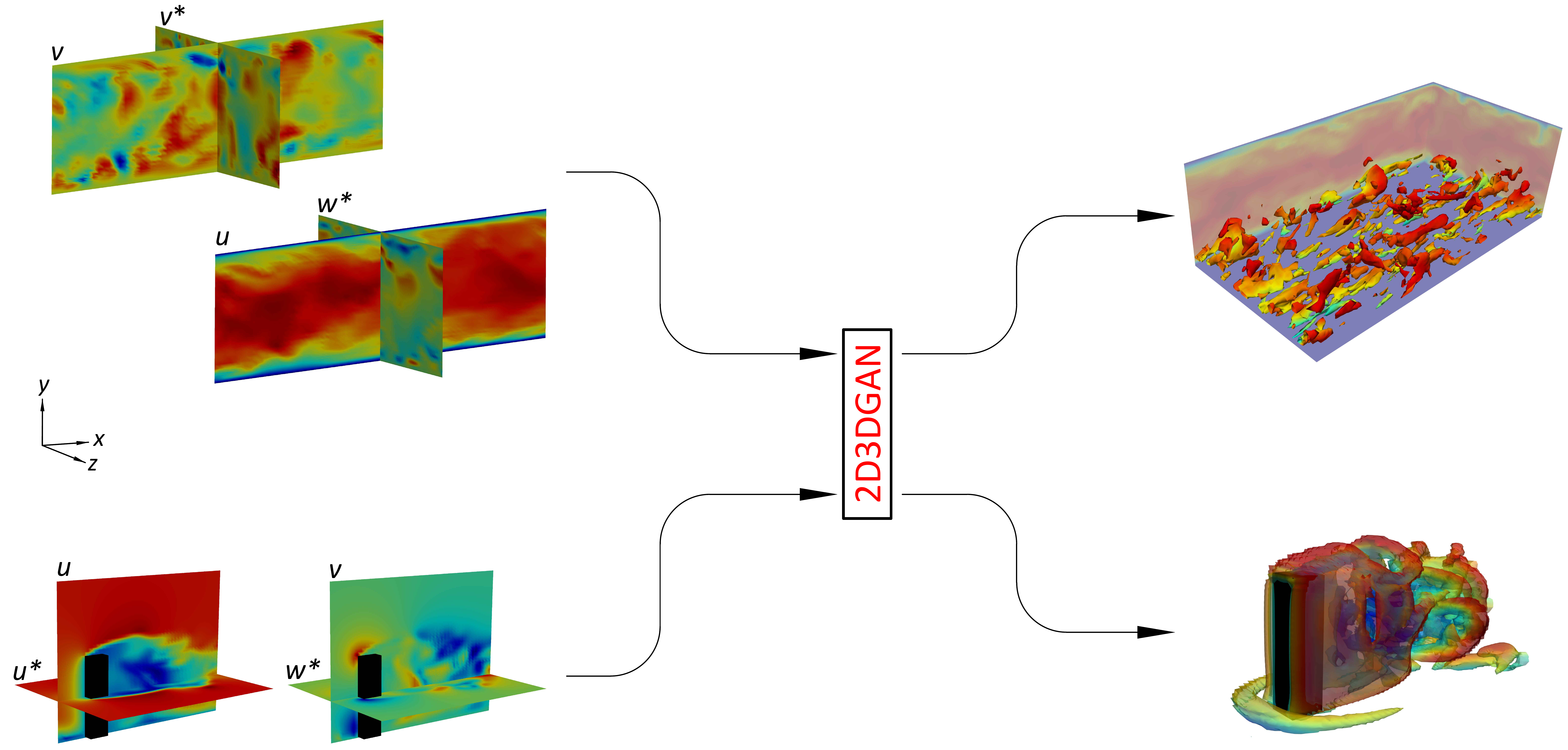}}
  \caption{\textbf{Procedure of reconstructing three-dimensional turbulent flows using the 2D3DGAN.} The domain from each simulation contains planar data of the flow obtained from a cross-plane. Note that the flow data of the two planes in the cross plane are entirely unpaired, $i.e.$ the velocity fields in the planes are at different instants. The superscript ‘$*$’ indicates the velocity components that are matched with the velocity data from the observation plane by applying the procedure explained in equation~(\ref{eqn:eq2}).}
\label{fig:F1}
\end{figure}

The proposed 2D3DGAN is inspired by the works of Wang {\it et al.} \citep{Wangetal2018} and Yousif {\it et al.} \citep{Yousifetal2021}, where the input to the network is data that contains limited information about the flow instead of the random vector $z$ in equation~(\ref{eqn:eq1}). While the models of \citep{Wangetal2018,Yousifetal2021} are designed to reconstruct 2D high-resolution data from spatially limited 2D data, our 2D3DGAN is designed to reconstruct 3D flow fields from 2D data. As shown in figure~\ref{fig:F1}, the input to the 2D3DGAN is data represented by a cross-plane of unpaired intersected 2D flow observation planes, $i.e.$ each plane contains data of two velocity components, and the data are collected at a period that is different from the other plane. Hence, the velocity fields in the planes are at different instants. \par

In this study, two cases are used to evaluate the performance of the proposed model: a turbulent channel flow at two different friction Reynolds number, $Re_\tau =$180 and 500, and the flow around a finite wall-mounted square cylinder with aspect ratio, $AR = $4, at a Reynolds number based on the free stream velocity and the cylinder width $d$ of $Re_d = 500$. \par

\begin{figure}
  \centerline{\includegraphics[scale=0.15]{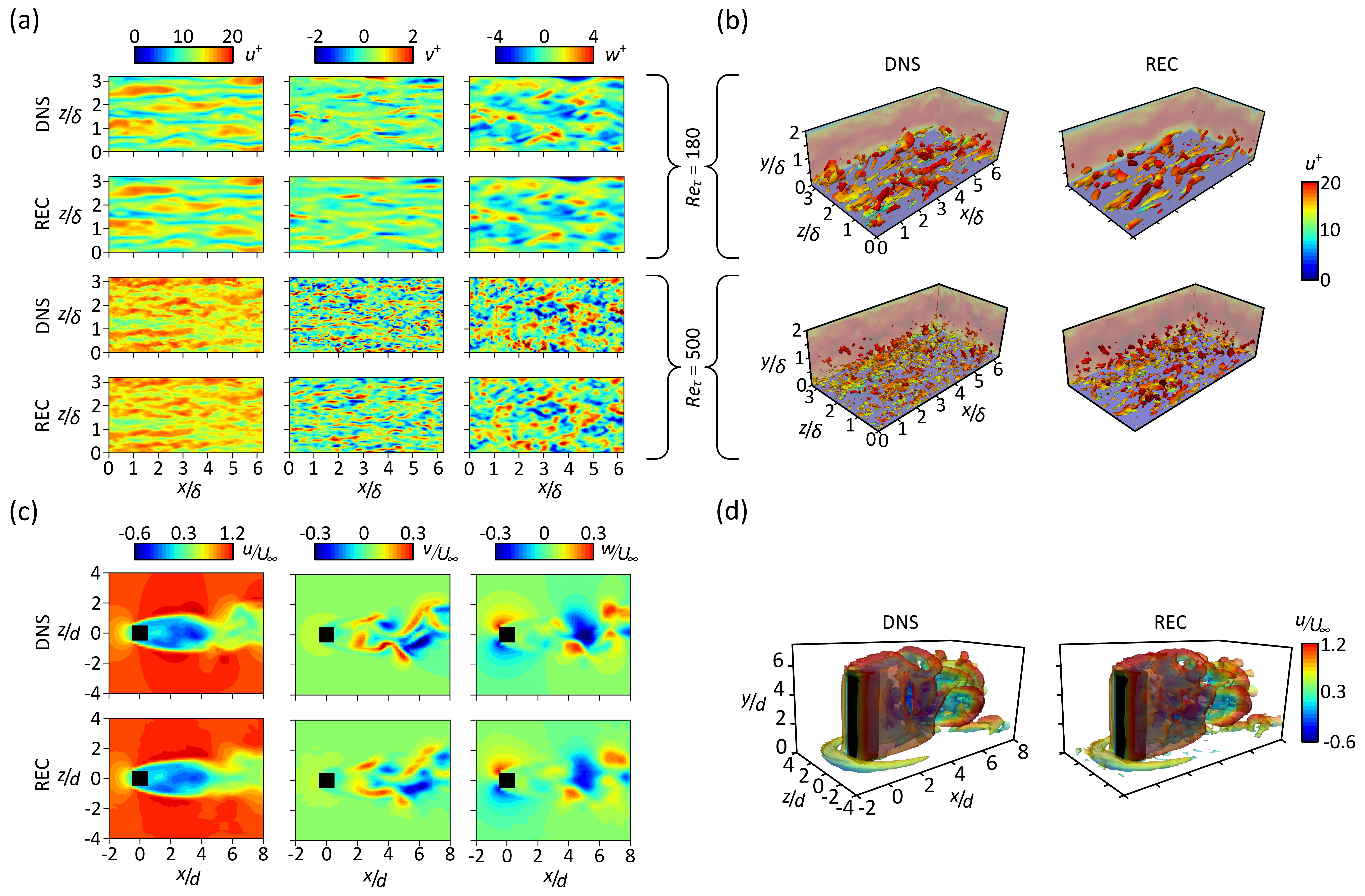}}
  \caption{\textbf{Instantaneous velocity fields and flow structures.} (a) $(x-z)$ Plane instantaneous velocity fields of the turbulent channel flow case; from left to right: streamwise velocity, wall-normal velocity and spanwise velocity; the planes are located at $y^+ = 16.78$ and 46.95, for the flow at $Re_\tau = 180$ and $500$, respectively. (b) Isosurfaces of the instantaneous flow structure for the turbulent channel flow case; $Q^+ = 0.006$ and 0.002 for the flow at $Re_\tau =$ 180 and 500, respectively. (c) Instantaneous velocity fields for the case of flow over a finite wall-mounted square cylinder at the $(x-z)$ plane, located at $y/d = 3$; from left to right: streamwise velocity, wall-normal velocity and spanwise velocity. (d) Isosurfaces of the instantaneous flow structure for the case of flow around a finite wall-mounted square cylinder; $Q (d/U_\infty )^2 = 0.0068$. Note that REC denotes reconstructed field.}
\label{fig:F2}
\end{figure}

Let us take the case of turbulent channel flow in figure~\ref{fig:F1} as a demonstration of matching the unpaired data before feeding it to the 2D3DGAN. The data from the two sections are synthetically unpaired such that no instantaneous velocity of each section is found at the same time as the instantaneous velocity in the other section. After that, the flow data of the ($x-y$) section is matched with the data of ($y-z$) section by utilising the square of $L_2$ norm error for the intersection line of the wall-normal velocity such that:

\begin{equation} \label{eqn:eq2}
v_{(y-z)}^*w_{(y-z)}^*= \substack{{\rm argmin}\\ v_{(y-z)} , w_{(y-z)}} \left( \left\| v_{(x-y)}^y - v_{(y-z)}^y (t) \right\|_2^2  \right),          
\end{equation}

\noindent where $v$ and $w$ are the wall-normal and spanwise instantaneous velocity components. Here, $v_{(x-y)}^y$ represents the wall-normal velocity in the $(x-y)$ plane, and $v_{(y-z)}^y (t)$ represents the wall-normal velocity in the $(y-z)$ plane as a function of time, $t$. The superscript ‘$*$’ indicates the matched velocity data using equation~(\ref{fig:F2}). This procedure mimics the matching of two data sets from two planar PIV experiments conducted each for a different plane. The first plane is the observation plane (which is here represented by the ($x-y$) plane) with velocity components $(u,v)$, where $u$ is the instantaneous streamwise velocity component, and the second one is the $(y-z)$ plane with velocity components $(v,w)$. Note that the 3D label data used for training the model are at the same instants of the observation plane. A similar approach is followed for the case of flow over a finite wall-mounted square cylinder with the observation plane being the central $(x-y)$ plane $(z/d = 0)$. The $(x-z)$ plane at $y/d =2$ is used to provide the matched velocity data with the data from the observation plane. \par

\vspace{0.5cm}
\noindent \textbf{Instantaneous flow reconstruction.} First, we examine the ability of the model to reconstruct the 3D instantaneous velocity fields. Figure ~\ref{fig:F2}(a) shows the reconstructed instantaneous velocity fields of the case of turbulent channel flow $(u^+ , v^+ , w^+ )$ at the $(x-z)$ plane, located at $y^+ = 16.78$ and 46.95, for the flow at $Re_\tau = 180$ and $500$, respectively. As can be seen in the figure, the velocity fields are successfully reconstructed by the model with commendable precision even though the $(x-z)$ plane is not introduced to the model during the training process. Note that the superscript ‘$+$’ indicates that the quantity is scaled in inner units using the fluid viscosity, $\nu$ and the friction velocity, $u_\tau$. The flow structure is investigated in figure~\ref{fig:F2}(b) by utilising the $Q$-criterion for vortex identification \citep{lozanoetal2014}. As shown in the figure, the reconstructed instantaneous velocity fields reveal a vortical structure $(Q^+)$ that is similar to that obtained from the DNS data, indicating that the model could successfully reproduce the flow fields with high accuracy. \par

Figure~\ref{fig:F2}(c) shows the reconstructed instantaneous velocity fields for the case of flow around a finite wall-mounted square cylinder $(u⁄U_\infty ,v⁄U_\infty ,w⁄U_\infty )$ at $y/d = 3$, where $U_\infty$ is the free stream velocity. As shown in the figure, all the three velocity fields are in good agreement with the DNS data. Note that the velocity fields along the height of the cylinder show very good agreement with the DNS data even for this region, $i.e.$ near the free end of the cylinder, which is not seen by the model as input during the training process. This indicates that the model can reconstruct the complex three-dimensional turbulent flow around the cylinder within all the regions. Furthermore, the vortical structure obtained from the reconstructed instantaneous velocity fields is generally consistent with the results from DNS as shown in figure~\ref{fig:F2}(d).

Finally, the accuracy of the reconstruction is examined via the $L_2$-norm relative error:

\begin{equation} \label{eqn:eq3}
\epsilon = \frac{1}{J} \sum_{j=1}^J  \frac{\left\| {\alpha}_j^{\rm DNS} - {\alpha}_j^{\rm REC}\right\|_2 }{\left\| {\alpha}_j^{\rm DNS}\right\|_2}, 
\end{equation}
where $\alpha_j^{\rm DNS}$ and $\alpha_j^{\rm REC}$  represent the ground truth (DNS) and the reconstructed instantaneous velocity components, respectively, and $J$ represents the number of the test snapshots. As can be observed from table~\ref{tab:Table1}, for the case of turbulent channel flow, no significant increase in the error values is observed in the channel when increasing $Re_{\tau}$ from 180 to 500, indicating the robustness of the model to increasing Reynolds number. Also, for the case of flow around a finite wall-mounted square cylinder, the error values are acceptable as compared to the error values of the turbulent channel, a fact that further supports the ability of the model to reconstruct the velocity fields of complex flows. Note that the errors in $u$ are low, while the errors in $v$ and $w$ are comparatively higher. This is explained by the fact that in these flows the main physics is driven by the streamwise component, which is the main focus of the deep learning model when performing the predictions. It is worth noting here that in GAN-based models, the mapping of high frequency fluctuations in the data is related to the adversarial loss. In other words, in GAN-based models, synthetic data is generated for the high frequency fluctuations. This is the main difference between GAN-based models and traditional CNNs, where the results are usually blurry with few flow details that can be predicted. Also, it is important to note that, despite the deviations in $v$ and $w$ error values, the main flow features are very well reproduced (as observed in figure~\ref{fig:F2}), and as discussed from the statistical and spectral perspectives next.

\begin{figure}
  \centerline{\includegraphics[scale=0.28]{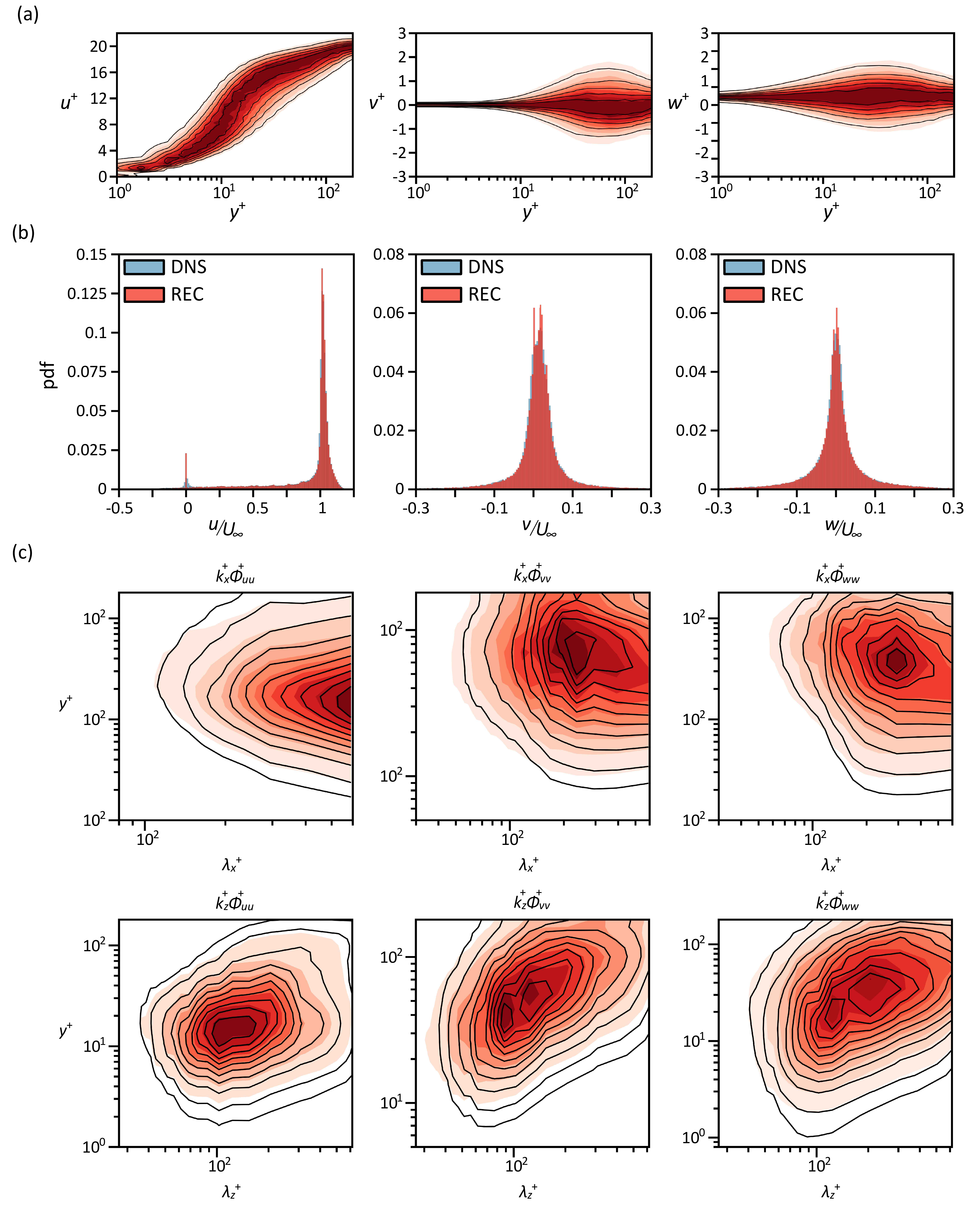}}
  \caption{\textbf{Probability density functions of the velocity components.} (a) pdf of the turbulent channel flow case as a function of $y^+$; from left to right: streamwise velocity, wall-normal velocity and spanwise velocity. The shaded contours represent the results from the DNS data and the lines represent the results from the reconstructed data. The contour levels are in the range of 20\%-80\% of the maximum pdf with an increment of 20\%. Results for the flow at $Re_\tau =$ 180. (b) pdf of the velocity components for the case of flow over a finite wall-mounted square cylinder; from left to right: streamwise velocity, wall-normal velocity and spanwise velocity. Note that REC denotes reconstructed field. (c) Premultiplied streamwise (top) and spanwise (bottom) power-spectral density of the velocity components from the turbulent channel flow case, as a function of $y^+$ and $\lambda^+$; from left to right: streamwise velocity, wall-normal velocity and spanwise velocity. The shaded contours represent the results from the DNS data and the lines represent the results from the reconstruction. The contour levels are in the range of 10\%-90\% of the maximum $ k_x^+ \mathit\Phi_{\alpha \alpha}^+ $ and $ k_z^+ \mathit\Phi_{\alpha \alpha}^+ $, with increments of 10\%. Results for the flow at $Re_\tau =$180.}
\label{fig:F3}
\end{figure}

\vspace{0.5cm}
\noindent \textbf{Spatial distribution of the flow.} In order to examine the ability of the proposed model to reconstruct the velocity fields with accurate spatial distribution, the probability density function of each velocity component, pdf is calculated for both cases. As shown in figure~\ref{fig:F3}(a), the pdf plots of the reconstructed velocity components for the case of turbulent channel flow show a remarkable agreement with those obtained from the DNS data, for all the regions along $y^+$. Also, for the case of flow over a finite wall-mounted square cylinder (figure~\ref{fig:F3}(b)), we can see that the pdf plots are consistent with the DNS data for all the three velocity components. Here, the results indicate the ability of the model to reconstruct the instantaneous velocity fields with accurate spatial distributions. \par

Furthermore, the ability of the model to reproduce the spectral content of the flow in the turbulent channel is investigated by utilising the premultiplied streamwise and spanwise power-spectral densities, $i.e.$ $ k_x \mathit\Phi_{\alpha \alpha} $ and $ k_z \mathit\Phi_{\alpha \alpha} $, where $k_x$ and $k_z$ are the streamwise and spanwise wavenumbers, respectively, and $\mathit\Phi_{\alpha \alpha}$ represents the corresponding wavenumber spectrum of each velocity component, $\alpha$. Figure~\ref{fig:F3}(c) shows $k_x^+ \mathit\Phi_{\alpha \alpha}^+$ and $k_z^+ \mathit\Phi_{\alpha \alpha}^+ $ as a function of $y^+$ and the corresponding inner-scaled wavelengths, $\lambda_x^+$ and $\lambda_z^+$. The spectra of the reconstructed velocity components are generally in agreement with the results obtained from DNS, with a slight deviation at high-wavenumber regions. These results further support the performance of the proposed model to properly represent the spatial distribution of the velocity fields. \par

\begin{table}
  \begin{center}
   \caption{$L_2$-norm relative error of the reconstructed velocity fields.}

\begin{tabular}{lccc}
\hline\\
Case&~~~$\epsilon(u)$&~~~~$\epsilon(v)$&~~~~$\epsilon(w)$\\\\
\hline\\

Turbulent channel flow $Re_{\tau}= 180$  &0.041&~~~0.577&~~~0.471\\
Turbulent channel flow $Re_{\tau}= 500$&0.044&~~~0.589&~~~0.515\\
Flow around a finite wall-mounted square cylinder&0.068&~~~0.743&~~~0.649\\\\
\hline\\
\end{tabular}
 
  \label{tab:Table1}
  \end{center}
\end{table}

\begin{figure}
  \centerline{\includegraphics[scale=0.18]{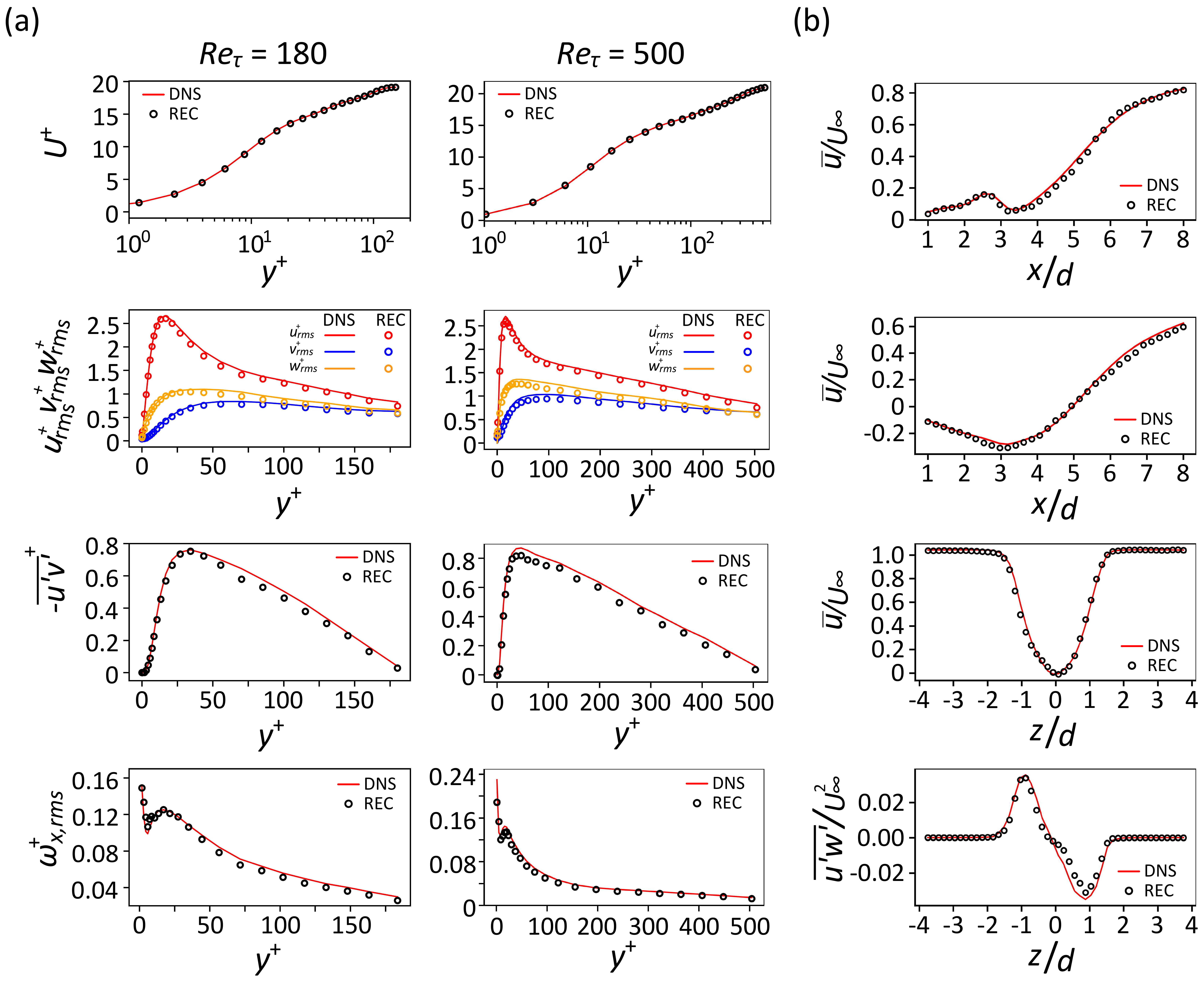}}
  \caption{\textbf{Turbulence statistics.} (a) Turbulence statistics of the turbulent channel flow case; from top to bottom: mean streamwise velocity profile, root-mean-square profiles of the velocity components, Reynolds shear-stress stress profile and root-mean-square profile of the streamwise vorticity. (b) Turbulence statistics of the flow around a finite wall-mounted square cylinder case; from top to bottom: streamwise profile of mean streamwise velocity at $y/d =1$, streamwise profile of mean streamwise velocity at $y/d = 3$, spanwise profile of mean streamwise velocity at $y/d = 3$ and spanwise profile of Reynolds  stress at $y/d =3$. Note that REC denotes reconstructed field.}
\label{fig:F4}
\end{figure}

\vspace{0.5cm}
\noindent \textbf{Turbulence statistics.} To examine the capability of the proposed model to reproduce the turbulence statistics, the first and second-order turbulence statistics of the reconstructed velocity fields are calculated and shown in figure~\ref{fig:F4}. As can be observed in figure~\ref{fig:F4}(a), for the case of turbulent channel flow at  $Re_\tau = 180$, the profile of the mean streamwise velocity $(U^+ )$ and the root-mean-square (rms) profiles of the velocity components $(u_{\rm rms}^+, v_{\rm rms}^+, w_{\rm rms}^+)$ are in excellent agreement with the results obtained from DNS. The Reynolds stress profile $(-\overline{u'v'}^+)$ exhibits a relatively good agreement with the results obtained from DNS, with a slight deviation in the range between $y^+$ = 50 and 125. The fluctuating streamwise vorticity $(\omega_{x, {\rm rms}}^+)$ profile is also in agreement with the results obtained from DNS. Here the results indicate that the model can reproduce the turbulence statistics of the flow with remarkable accuracy. The turbulence statistics for the flow at $Re_\tau = 500$ indicate that the reconstructed data are generally in good agreement with the DNS. Nonetheless, the Reynolds shear-stress profile exhibits a slight underprediction for a region that starts from the maximum shear stress and continues along $y^+$.
As can be seen in figure~\ref{fig:F4}(b), for the case of the flow over a wall-mounted square cylinder, the streamwise, as well as the spanwise profile of the mean streamwise velocity $(\overline{u}⁄U_\infty )$ exhibit an excellent agreement with the results obtained from the DNS for the examined elevations along the cylinder height. The comparison of the spanwise profile of the Reynolds shear stress $(\overline{u'w'}⁄{U_\infty^2})$ also reveals generally good agreement with the results obtained from the DNS. However, it exhibits a slight deviation at the examined wall-normal location, $i.e.$ $y/d = 3$. This might be attributed to the increase in the complexity of the flow as the wall-normal distance increases \citep{Bourgeoisetal2011, Saha2013, Yousif&Lim2021, Lazpitaetal2022}.

\section*{Discussion and conclusions}\label{sec3}

This study presents a deep-learning-based method to reconstruct 3D turbulent flows from 2D data. A combination of 2D and 3D CNNs was utilised to build a model based on GAN, named 2D3DGAN. The model is designed to reconstruct 3D turbulent flows from a cross-plane of unpaired data (the data of each plane in the cross-plane is collected at a different time interval), such that it can be utilised to reconstruct 3D turbulent velocity fields from two-dimensional PIV measurements. \par

The reconstructed instantaneous velocity fields and the 3D flow structures show an excellent agreement with the DNS results for both the cases used to test the model, $i.e.$ turbulent channel flow and the flow around a finite wall-mounted square cylinder, even for the regions where the data is not introduced to the model during the training process. The error analysis reveals that the model is robust to increasing Reynolds number and the complexity of the flow. The model also successfully reproduces the turbulence statistics with very good accuracy. Furthermore, the flow spectra for the case of turbulent channel flow also reveal a commendable agreement with the result obtained from the DNS data, indicating a great ability of the model to maintain a realistic spatial behaviour of the velocity fields. \par

This study demonstrates for the first time that GAN-based models can be successfully used for reconstructing 3D turbulent flows from 2D data. This approach opens the doors to discovering new data-driven methods that can reconstruct the 3D turbulent flows from 2D experimental measurements, which can result in a remarkable cost reduction in complexity of the experimental setup required to achieve accurate 3D experimental measurements. \par

\section*{Methods}\label{sec4}

\begin{figure}
  \centerline{\includegraphics[scale=0.16]{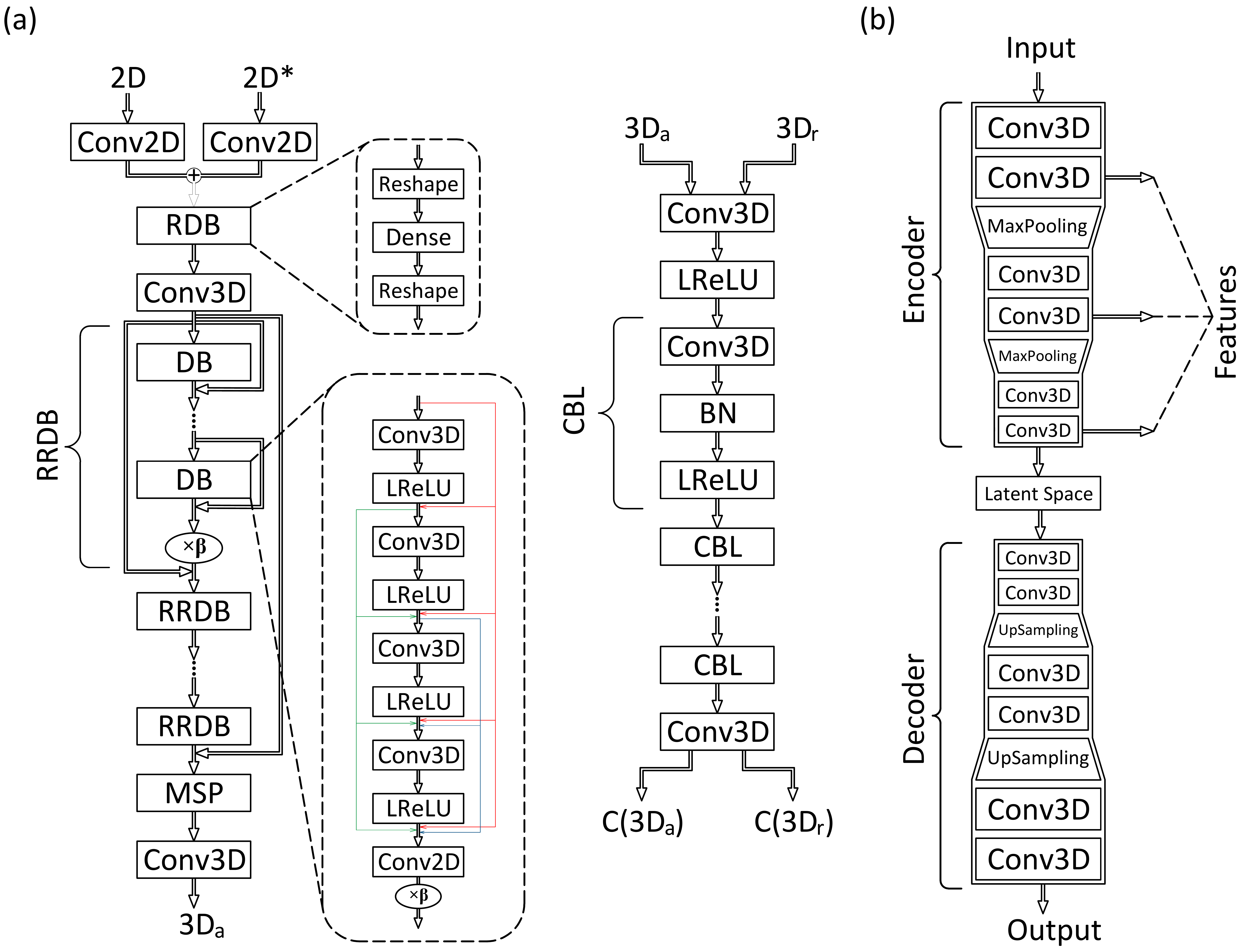}}
  \caption{\textbf{Architecture of  2D3DGAN and 3DCAE.} (a) Schematic of 2D3DGAN showing the generator network (left) and the discriminator network (right). The generator network receives a dataset represented by two planes in the flow, the first one is the observation plane, and the second one is the plane that is matched with the observation plane by applying the procedure explained in equation~(\ref{eqn:eq2}), which is indicated by the superscript ‘$*$’. The output from the generator network, $i.e.$ the artificial 3D flow data is fed to the discriminator network, and the latter tries to distinguish if the data is artificial or true. (b) Schematic of the feature extractor (3DCAE). The main features of the flow fields are extracted using three layers in the encoder part of the 3DCAE.}
\label{fig:F5}
\end{figure}

\textbf{Building the 2D3DGAN architecture.} The architecture of 2D3DGAN is shown in figure~\ref{fig:F5}(a). Here, $G$ consists of two stages. First, two sections of 2D data are introduced to $G$ and passed through a 2D CNN and two stages of reshaping with a dense layer between them. After that, the 3D data are passed through a deep 3D CNN represented by residual in residual dense blocks (RRDBs) and multi-scale part (MSP). The MSP, which consists of three parallel 3D convolutional sub-models with different kernel sizes, is applied to the data features extracted by the RRDBs. More details regarding MSP can be found in Yousif {\it et al.} \citep{Yousifetal2021, Yousifetal2022b}. The outputs of the three sub-models are summed and passed through a final 3D convolutional layer to generate an artificial 3D data ($3D_a$). The artificial and real data are fed to $D$ and passed through a series of 3D convolutional, batch normalization, and leaky-ReLU layers. As a final step, the data are passed through a 3D convolutional layer. The non-transformed discriminator outputs using the real and artificial data, $i.e.$ $C(3D_r )$ and $C(3D_a )$, are used to calculate the relativistic average discriminator value ($D_{Ra}$) \citep{Jolicoeur-Martineau2018}:

\begin{equation} \label{eqn:eq4}
D_{Ra} (3D_r , 3D_a ) = \sigma (C \left(3D_r ) \right) - \mathbb{E}_{3D_a} \left[C ( 3D_a ) \right],
\end{equation}

\begin{equation} \label{eqn:eq5}
D_{Ra} (3D_a , 3D_r ) = \sigma (C \left(3D_a) \right) - \mathbb{E}_{3D_r} \left[C (3D_r ) \right],
\end{equation}

\noindent where $\sigma$ is the sigmoid function. In Equations~(\ref{eqn:eq4}) and ~(\ref{eqn:eq5}), $D_{Ra}$ represents the probability that the output from $D$ using the real 3D data is relatively more realistic than the output using the generated 3D data. \par

The discriminator loss is then defined as:

\begin{equation} \label{eqn:eq6}
L_D^{Ra} = -\mathbb{E}_{3D_r} \left[ {\rm log} (D_{Ra} (3D_r , 3D_a)) \right] - \mathbb{E}_{3D_a} \left[ {\rm log} (1 - D_{Ra} (3D_a, 3D_r )) \right].
\end{equation}

The adversarial loss for the generator can be expressed in a symmetrical form as:

\begin{equation} \label{eqn:eq7}
L_G^{Ra} = -\mathbb{E}_{3D_r} \left[ {\rm log} (1 - D_{Ra} (3D_r , 3D_a )) \right] - \mathbb{E}_{3D_a} \left[ {\rm log} (D_{Ra} (3D_a , 3D_r )) \right].
\end{equation}

The total loss function of $G$ is defined as:

\begin{equation} \label{eqn:eq8}
\mathcal{L}_G =  \beta_1 L_G^{Ra} + \beta_2 L_{\rm voxel} + \beta_3 L_{\rm perceptual} + L_{\rm continuity} + L_{\rm momentum},     
\end{equation}

\noindent where $L_{\rm voxel}$ is the error calculated based on the voxel (volume pixel) difference between the generated data and the ground truth data. $L_{\rm perceptual}$ represents the difference in the extracted features of the real and the generated data. In this study, we use a 3D convolutional auto-encoder (3DCAE) to extract the features from the data obtained from the generator and the ground truth data as shown in figure~\ref{fig:F5}(b). Note that $L_{\rm continuity}$ and $L_{\rm momentum}$ represent the error of the continuity and momentum equations. In the loss function, $\beta_1$, $\beta_2$, and $\beta_3$ are the coefficients used to balance the loss terms, and their values are set to 10, 1000 and 2000, respectively. The square of the $L_2$ norm error is used to calculate all the loss terms of the generator except $L_G^{Ra}$, such that:

\begin{equation} \label{eqn:eq9}
L_{\rm voxel}= \frac{1}{M} \sum_{m=1}^M \left\| 3D_r - 3D_a \right\|_2^2,           
\end{equation}

\begin{equation} \label{eqn:eq10}
L_{\rm perceptual}= \frac{1}{M} \sum_{m=1}^M \left\| \mathscr{F}_{FE} (3D_r )- \mathscr{F}_{FE} (3D_a ) \right\|_2^2,            
\end{equation}

\begin{dmath} \label{eqn:eq11}
L_{\rm continuity}  = \frac{1}{M} \sum_{m=1}^M \left\| \left( \frac{\partial u_r }{\partial x} + \frac{\partial v_r }{\partial y} + \frac{\partial w_r }{\partial z} \right) - \left( \frac{\partial u_a }{\partial x} + \frac{\partial v_a }{\partial y} + \frac{\partial w_a }{\partial z} \right) \right\| _2^2,    
\end{dmath}

\begin{dmath} \label{eqn:eq12}
L_{\rm momentum} = L_{\rm momentum}^u + L_{\rm momentum}^v + L_{\rm momentum}^w,  
\end{dmath}

\begin{dmath} \label{eqn:eq13}
L_{\rm momentum}^u = \frac{1}{M} \sum_{m=1}^M \left\| \left( \frac{\partial u_r }{\partial t} + u_r \frac{\partial u_r }{\partial x} + v_r \frac{\partial u_r }{\partial y} + w_r \frac{\partial u_r }{\partial z} \right) - \left( \frac{\partial u_a }{\partial t} + u_a \frac{\partial u_a }{\partial x} + v_a \frac{\partial u_a }{\partial y} + w_a \frac{\partial u_a }{\partial z} \right)  \right\|_2^2, \end{dmath}

\begin{dmath} \label{eqn:eq14}
L_{\rm momentum}^v = \frac{1}{M} \sum_{m=1}^M \left\| \left( \frac{\partial v_r }{\partial t} + u_r \frac{\partial v_r }{\partial x} + v_r \frac{\partial v_r }{\partial y} + w_r \frac{\partial v_r }{\partial z} \right) - \left( \frac{\partial v_a }{\partial t} + u_a \frac{\partial v_a }{\partial x} + v_a \frac{\partial v_a }{\partial y} + w_a \frac{\partial v_a }{\partial z} \right)  \right\|_2^2, \end{dmath}

\begin{dmath} \label{eqn:eq15}
L_{\rm momentum}^w = \frac{1}{M} \sum_{m=1}^M \left\| \left( \frac{\partial w_r }{\partial t} + u_r \frac{\partial w_r }{\partial x} + v_r \frac{\partial w_r }{\partial y} + w_r \frac{\partial w_r }{\partial z} \right) - \left( \frac{\partial w_a }{\partial t} + u_a \frac{\partial w_a }{\partial x} + v_a \frac{\partial w_a }{\partial y} + w_a \frac{\partial w_a }{\partial z} \right)  \right\|_2^2, \end{dmath}

\noindent where $M$ represents the number of snapshots in the training mini-batch, which is fixed in this study to 8, and $\mathscr{F}_{FE}$  in equation~(\ref{eqn:eq10}) represents the function of the feature extractor, $i.e.$ the 3DCAE. 

\vspace{0.5cm}
\noindent \textbf{Overview of the training setup.} The 2D3DGAN is trained separately with the data of turbulent channel flow at $Re_\tau = u_\tau \delta ⁄ \nu = 180$ and 500, and with the data of flow around a finite wall-mounted square cylinder with $AR = $4, and  at $Re_d = U_\infty d⁄\nu = 500$, where, $\delta$ is the channel half-width. For the case of turbulent channel flow, the transfer-learning technique \citep{Guastonietal2021, Yousifetal2021, Yousifetal2022a} is used to train the model for the flow at $Re_\tau = 500$, in order to further reduce the computational cost (represented by the training time) and the required training data. For both cases, 80\% of the collected data are used as training set and the rest, $i.e.$ 20\%, are used as testing set. The open-source library TensorFlow 2.4.0 \citep{Abadietal2016} is used for the implementation of the 2D3DGAN. The adaptive moment estimation (Adam) optimization algorithm \citep{Kingma&Ba2017} is used to update the weights of the model. The learning rate is initially set to $10^{-4}$ and progressively decreased  during the training process to finally reach $1.25 \times 10^{-5}$. The number of trainable parameters in the model is approximately 5.7 million for the generator network and 3.74 million for the discriminator. The training of the model using a machine with a single NVIDIA TITAN RTX GPU for the case of turbulent channel flow requires approximately 48 hours for $Re_\tau = 180$ and for $Re_\tau = 500$, with the aid of transfer learning, requires approximately 30 hours. The training of the case of flow over a wall-mounted square cylinder requires approximately 40 hours.
 
\vspace{0.5cm}
\noindent \textbf{Direct numerical simulation of turbulent channel flow.} DNS calculations of a fully-developed incompressible turbulent channel flow at friction Reynolds numbers $Re_\tau = 180$ and 500 are performed to generate the training and testing datasets. The incompressible Navier--stokes equations have been solved using the LISO code \citep{Hoyas&Jimenez2006}, similar to the one described by Lluesma-Rodriguez {\it et al.}~\citep{Lluesmaetal2021}. This code has successfully been employed to run some of the largest simulations of wall-bounded turbulent flows \citep{alcantara2021, Hoyasetal2022, Oberlacketal2022}. Briefly, the code uses the same strategy as that described by Kim {\it et al.}~\citep{kimetal1987}, but uses a seven-point compact-finite difference scheme in the $y$ direction with fourth-order consistency and extended spectral-like resolution \citep{Lele1992}. The temporal discretization is a third-order semi-implicit Runge–Kutta scheme \citep{Spalartetal1991}. The wall-normal grid spacing is adjusted to keep the resolution to $\Delta y = 1.5 \eta$, $i.e.$ approximately constant in terms of the local isotropic Kolmogorov scale $\eta = \left( \nu^3 / \varepsilon \right)^{0.25}$, where $\varepsilon$ is the isotropic dissipation of turbulent kinetic energy.

The dimensions of the computational domain for both simulations are set to $8 \pi \delta, 2 \delta$ and $3 \pi \delta$ in the streamwise, wall-normal and spanwise directions, respectively. A uniform distribution of the grid points is used in the streamwise and spanwise directions and a non-uniform distribution is used in the non-homogeneous wall-normal direction. For the flow at $Re_\tau = 180$, the grid spacings in the streamwise ($\Delta x^+ $) and spanwise directions ($\Delta z^+$) are 8.55 and 4.27, respectively, while for the flow at  $Re_\tau =$ 500, $\Delta x^+ = 8.33 $ and $\Delta z^+ = 4.16 $. The grid spacing near the wall in the wall-normal direction is $\Delta y_w^+ = 0.53$ and $0.74$ for the flow at $Re_\tau =$ 180 and 500, respectively. The simulation time step ($\Delta t^+$) is set to 0.07 and 0.09 for the flow at $Re_\tau =$ 180 and 500, respectively. The flow is periodic in the streamwise and spanwise directions, whereas the no-slip condition is applied to the channel walls. A total of 1,000 consecutive snapshots are collected for each of the two simulations. \par

\vspace{0.5cm}
\noindent \textbf{Direct numerical simulation of flow around a finite wall-mounted square cylinder.} In the second case, we consider the flow around a finite wall-mounted square cylinder with $AR = $4, at $ Re_d = 500$. The spectral-element-method (SEM)-based open-source code Nek5000 developed by Fischer {\it et al.} \citep{Fischeretal2022} is used to perform the DNS. In the SEM \citep{Patera1984}, the computational domain is decomposed into elements, and the solution is expressed in terms of Lagrange interpolants of order $N$ within those elements. The location of the nodes inside the elements follows the Gauss--Lobatto--Legendre (GLL) distribution, whereas there is an isoparametric mapping for the shape of the elements and there are no restrictions regarding the position of the elements in the domain. This means that this method allows the flexibility to compute complex geometries, while still preserving the characteristics of a high-order spectral method. In the present study, the velocity field is expressed in terms of Lagrange interpolants of order $N=5$, and order $N-2=3$ is considered for the pressure field. The nonlinear terms are treated explicitly by third-order extrapolation (EXT3), whereas the viscous terms are treated implicitly by a third-order backward differentiation scheme (BDF3). The no-slip boundary condition is applied for the cylinder walls and the ground, while periodic boundary conditions are used in the spanwise direction. At the top of the domain, we impose a constant streamwise velocity, a zero spanwise velocity and zero stress in $y$. Furthermore, the input is a laminar boundary layer, and the output is the stabilized boundary condition by Dong {\it et al.} \citep{Dongetal2014}. The dimensions of the simulation domain are $( L_x, L_y, L_z )/d = (60, 12, 12) $ in $x$, $y$ and $z$, respectively, and the total number of grid points is around 20 million. We collect a total of 10,000 snapshots with a time setp among snapshots of $\Delta t U_\infty / d = 0.02$. \par

\vspace{0.5cm}
\noindent \textbf{Data preparation and pre-processing.}
To reduce the computational cost of training the model and to increase the training and testing data, each simulation domain in the case of turbulent channel flow is divided into 12 identical sub-domains having a size of $2 \pi \delta$, $2 \delta $ and $\pi \delta$. Furthermore, the grid size of each sub-domain is interpolated and reduced to $64 \times 48 \times 48$ grid points with a uniform distribution in the $x$ and $z$ directions and a non-uniform distribution in the $y$ direction. For the case of flow around a wall-mounted square cylinder, the size of the domain that is used for training and testing the model is set to $10d \times 7.6d \times 8d$, and the data are interpolated and reduced into a uniform grid distribution of $48 \times 48 \times 48$ points. The input data to the model are normalised using the min-max normalisation to obtain values between 0 and 1.

\section*{Data availability}\label{sec6}

All the data analysed in this paper were produced with the in-house and open-source softwares described in the code-availability statement. Reference data and the scripts used to produce the data figures, as well as instructions to train and test the 2D3DGAN are available on the following web page:

(\textbf{\url{https://fluids.pusan.ac.kr/fluids/65416/subview.do}}).

\section*{Code availability}\label{sec7}

The DNS calculations of the turbulent channel flow case are performed using in-house code LISO, contact SH for the code availability. The DNS calculations of the case of flow over a finite wall-mounted square cylinder are performed using the open-source code Nek5000: (\textbf{\url{https://nek5000.mcs.anl.gov/}}). The open-source library TensorFlow 2.4.0  is used for the implementation of the 2D3DGAN: (\textbf{\url{https://www.tensorflow.org/}}).

\section*{Acknowledgments}\label{sec8}

This work was supported by `Human Resources Program in Energy Technology' of the Korea Institute of Energy Technology Evaluation and Planning (KETEP), granted financial resource from the Ministry of Trade, Industry \& Energy, Republic of Korea (no. 20214000000140). In addition, this work was supported by the National Research Foundation of Korea (NRF) grant funded by the Korea government (MSIP) (no. 2019R1I1A3A01058576). This work was also supported by the National Supercomputing Center with supercomputing resources including technical support (KSC-2021-CRE-0244). R.V. acknowledges the financial support from the ERC Grant No. “2021-CoG-101043998, DEEPCONTROL”. SH was funded by Contract No. PID2021-128676OB-I00 of Ministerio de Ciencia, innovación y Universidades/ FEDER.

\section*{Author contributions}\label{sec9}

M.Z.Y. and H.L. designed the research. M.Z.Y. and L.Y. wrote the code of the deep-learning model with input from R.V. R.V. and S.H. performed the simulations and provided the data. M.Z.Y., L.Y. and H.L. wrote the manuscript. R.V. and S.H. reviewed and edited the manuscript. H.L. supervised the project.

\section*{Competing interests}\label{sec10}

The authors declare no competing interests.

\bibliography{nature2022}


\end{document}